\documentclass[11pt]{article}

\usepackage{epsfig,subfigure}
\usepackage{amssymb, amsmath}
\usepackage{color}
\def\QED{\mbox{\rule[0pt]{1.5ex}{1.5ex}}}

\usepackage{graphicx}

\usepackage{amssymb}
\usepackage{amsmath,amsfonts,amssymb}
\usepackage{verbatim}
\usepackage[bookmarks=false]{}
\usepackage[font={small}]{caption}

\newtheorem{theorem}{Theorem}

\newtheorem{lemma}{Lemma}

\newtheorem{ex}{\textbf{Example}}

\newcommand\blfootnote[1]{%
  \begingroup
  \renewcommand\thefootnote{}\footnote{#1}%
  \addtocounter{footnote}{-1}%
  \endgroup
}

\newcommand\numberthis{\addtocounter{equation}{1}\tag{\theequation}}

\begin{document}
\date{}
\title{On the Optimality of Treating Interference as Noise}
\author{ \normalsize Chunhua Geng, Navid Naderializadeh, A. Salman Avestimehr, and Syed A. Jafar  \\
}

\maketitle
   \blfootnote{Chunhua Geng (email: chunhug@uci.edu) and Syed A. Jafar (email: syed@uci.edu) are with the Center of Pervasive Communications and Computing (CPCC) in the Department of Electrical Engineering and Computer Science (EECS) at the University of California Irvine. Navid Naderializadeh (email: nn245@cornell.edu) and Salman Avestimehr (email: avestimehr@ece.cornell.edu) are with the Department of Electrical and Computer Engineering (ECE) at Cornell University.}

\begin{abstract}
It is shown that in  the $K$-user  interference channel, if for each user the desired signal strength is no less than the sum of the strengths of the strongest interference from this user and the strongest interference to this user (all values in dB scale), then the simple scheme of using point to point Gaussian codebooks with appropriate power levels at each transmitter and treating interference as noise at every receiver (in short, TIN scheme) achieves all points in the capacity region to within a constant gap. The generalized degrees of freedom (GDoF) region under this condition is a polyhedron, which is shown to be fully achieved by the same scheme, without the need for time-sharing. The results are proved by first deriving a polyhedral relaxation of the GDoF region achieved by TIN, then providing a dual characterization of this polyhedral region via the use of potential functions, and finally proving the optimality of this region in the desired regime. 
\end{abstract}


\section{Introduction}

Treating interference as noise (TIN) when it is  sufficiently weak is one of the key principles of interference management. As a robust principle that is also known to be  optimal  under certain conditions, TIN is  interesting both from practical and theoretical perspectives. 

From a practical perspective, TIN is attractive for its low complexity and robustness to channel uncertainty. TIN involves the use of only point-to-point channel codes, that are well understood, quite practical, and near-optimal in their ability to deal with unstructured noise. Further, since it requires only a coarse knowledge of the signal to interference and noise power ratio (SINR) at the transmitters, the overhead associated with acquiring channel state information at the transmitters (CSIT) is minimal for the TIN scheme. The practical appeal of the TIN scheme has motivated several studies of the achievable rate region of TIN in the literature. However, as noted by e.g., \cite{Charafeddine_TIN,Cheewei_sumrate}, in spite of the simplicity of TIN,  the structure of the TIN rate region is non-trivial --- it involves the optimization of the  power levels at the transmitters, and  is generally non-convex by itself, i.e., if time-sharing is not involved.

\begin{figure}[h]
\begin{center}
 \includegraphics[width= 8 cm]{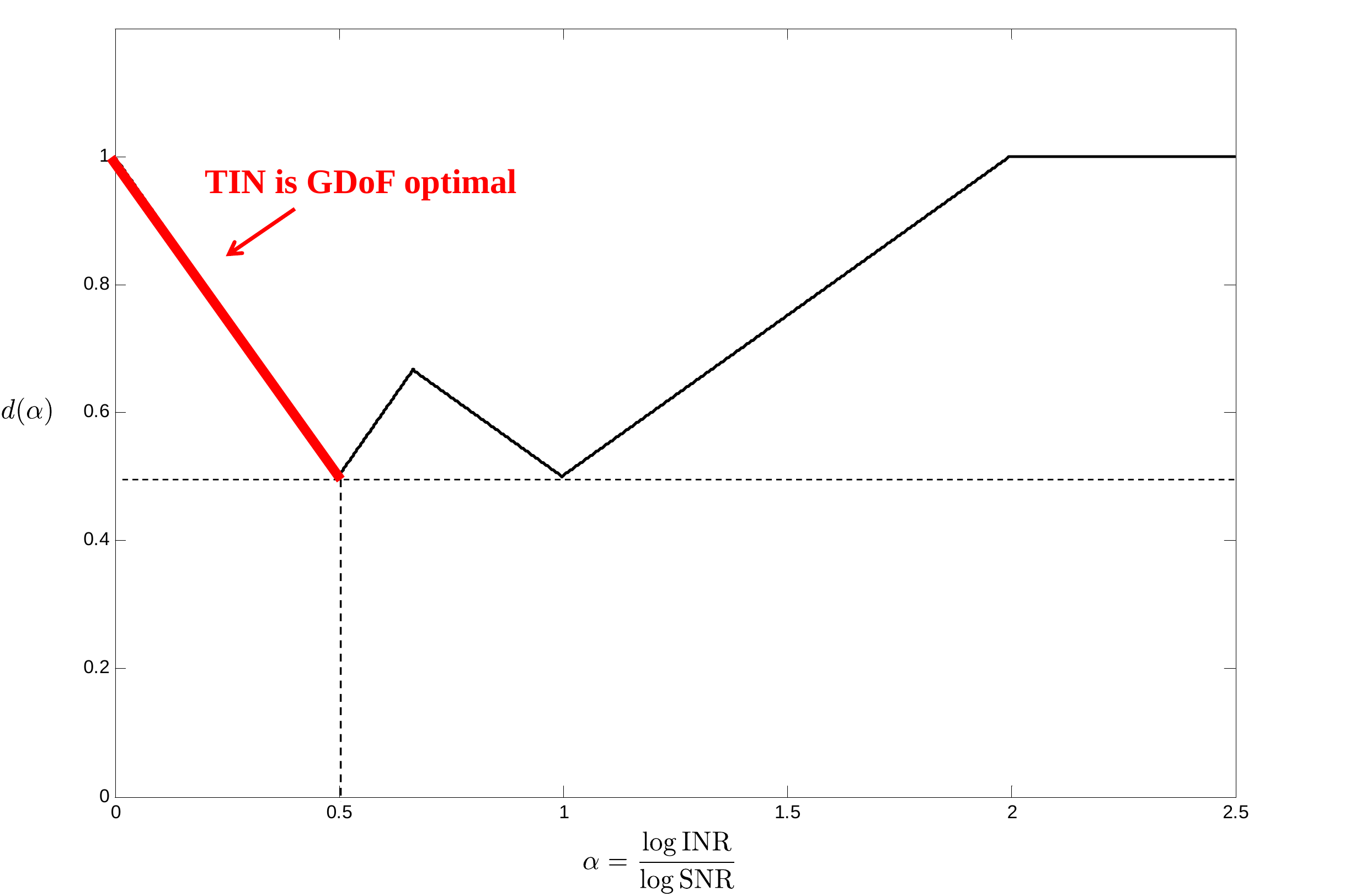}
 \caption{The GDoF ``W'' curve for the two-user symmetric Gaussian interference channel.}
\label{W_curve}
\end{center}
\end{figure}

From a theoretical perspective, it is the optimality of TIN that has attracted the most attention.  It is shown in  \cite{Motahari_Khandani_TIN,Annapureddy_Veeravalli_TIN_opt,Kramer_TIN_opt} that in a so-called ``noisy interference'' regime, TIN achieves the \emph{sum} capacity of the interference channel. An extension of the noisy interference regime is  obtained for multiple-input multiple-output (MIMO) Gaussian interference channels in \cite{Annapureddy_Veeravalli_MIMO}. In terms of generalized degree-of-freedom (GDoF), the well-known ``W'' curve  \cite{Tse_GDoF} in Fig. \ref{W_curve} demonstrates that for the two-user symmetric Gaussian interference channel, when the strengths of both direct channels are assumed as SNR and the interference channels are not stronger than $\sqrt{\mbox{SNR}}$, TIN  achieves the symmetric GDoF for each user. This result is  generalized to  $K$-user fully-connected symmetric Gaussian interference channels in \cite{Jafar_Vishwanath_GDoF} and to cyclic asymmetric Gaussian interference channels in \cite{Yu_Cyclic}. However,  not much is known about the regime where TIN is GDoF-optimal for the general \emph{fully-connected, fully-asymmetric} $K$-user  Gaussian interference channel.

In this work we present a general condition for the $K$-user fully-connected fully-asymmetric Gaussian interference channel, under which TIN is shown to be not only optimal for the entire GDoF region, but  also within a constant gap of the entire capacity region. The general  condition is stated in words as follows:

\emph{If for each user, the desired signal strength  is no less than the sum of the strengths of the strongest interference {from} this user and the strongest interference {to} this user (all values in dB scale), then TIN is GDoF optimal.}

As an example, consider a $K$-user interference channel, where all desired channels have strength SNR and the interference channels, each of which can  have a different strength, are not stronger than $\sqrt{\mbox{SNR}}$. In other words, the ratio of signal strength to the desired signal strength (each in dB scale) for each interferer is 0.5 or less and for the desired signal is 1. Then it follows, from the result shown in this work, that the GDoF-optimal scheme is TIN with appropriate power allocation at each transmitter.

Our proof of the optimality of TIN in the aforementioned regime consists of three steps. First, in this regime, we introduce a relaxed version of TIN, called \emph{polyhedral TIN}. Second, we show that, quite interestingly, the achievable GDoF region through polyhedral TIN, namely polyhedral TIN region, can be characterized by checking the existence of a potential function for an induced fully-connected directed graph, with nodes representing the source-destination pairs in the original interference channel (with the addition of a ``ground'' node), and a specific assignment of lengths to each arc. Using this equivalence and a potential theorem, we derive a dual characterization of the polyhedral TIN region. The significance of our dual characterization is the elimination of power allocation variables in the characterization. Finally, we use the outer bounds developed in~\cite{Yu_Cyclic} to establish the optimality of polyhedral TIN in the regime of interest. Since TIN performs no worse than polyhedral TIN, this proves that TIN is optimal in that regime in the GDoF sense.

Moreover, following the proof for GDoF-optimality of TIN, we show that in the same regime, TIN can also achieve the whole capacity region of the $K$-user interference channel to within a constant gap. Finally, we show that for general channel gain values in a $K$-user interference channel, the achievable GDoF region of TIN, namely TIN region, is composed of the union of $2^{K}-1$ polyhedra. However, for the regime of interest, one polyhedron subsumes all the others, hence the TIN region reduces to a single polyhedron, which is the polyhedral TIN region.

\section{System Model and Preliminaries}
As our starting point, consider the canonical model of a fully-asymmetric $K$-user wireless interference channel, with the input-output relationship
\begin{equation}
\label{ori_channel}
Y_k(t)=\sum_{i=1}^Kh_{ki}\tilde{X}_i(t)+Z_k(t),~~~\forall k\in\{1,2,...,K\},
\end{equation}
where at each time index $t$, $\tilde{X}_i(t)$ is the transmitted symbol of transmitter $i$, $Y_k(t)$ is the received signal of receiver $k$, $h_{ki}$ is the complex channel gain value from transmitter $i$ to receiver $k$, and $Z_k(t)\sim \mathcal{CN}(0,1)$ is the additive white Gaussian noise (AWGN) at receiver $k$. All the symbols are complex. Each transmitter $i$ is subject to the power constraint $E[|\tilde{X}_i(t)|^2]\leq P_i$.

We will translate the standard  channel model (\ref{ori_channel}) into an equivalent normalized form that is more conducive for GDoF studies. We define the signal-to-noise ratio (SNR) of user $i$ and interference-to-noise ratio (INR) of transmitter $i$ at receiver $k$ as follows\footnote{It is not difficult to verify that assigning a value of 1 to SNR's and INR's that are less than 1, or equivalently, assigning a 0 value to  $\alpha_{ij}$ that might otherwise be negative, is only a matter of convenience, and has no impact on the GDoF or the constant gap result.}.
\begin{equation}
\label{SNR_INR_Def}
\mathrm{SNR}_{i}\triangleq\max(1,|h_{ii}|^2P_i),~~~
\mathrm{INR}_{ki}\triangleq\max(1,|h_{ki}|^2P_i),~~~
i\neq k,~~i,k\in\{1,2,...,K\}.
\end{equation}

As in \cite{Tse_GDoF}, for the GDoF metric,  we preserve the ratios of different signal strengths in dB scale as all SNRs approach infinity. To this end, taking $P>1$ as a nominal power value, we define
\begin{equation}
\label{GDoF_Def}
\alpha_{ii}\triangleq\frac{\log \mathrm{SNR}_{i}}{\log P},~~~
\alpha_{ki}\triangleq\frac{\log \mathrm{INR}_{ki}}{\log P},~~~
i\neq k,~~i,k\in\{1,2,...,K\},
\end{equation}
implying that for each user $i$, $\mathrm{SNR}_i=P^{\alpha_{ii}}$ and for any two distinct users $i,k$, $\mathrm{INR}_{ki}=P^{\alpha_{ki}}$. 

Now according to (\ref{SNR_INR_Def}) and (\ref{GDoF_Def}), we can represent the original channel model in (\ref{ori_channel}) in the following  form,
\begin{equation}
\label{equ_channel}
\begin{aligned}
Y_k(t)=\sum_{i=1}^K\sqrt{P^{\alpha_{ki}}}e^{j\theta_{ki}}X_i(t)+Z_k(t),~~~ \forall k\in\{1,2,...,K\}.
\end{aligned}
\end{equation}

In this equivalent channel model, $X_i(t)=\tilde{X}_i(t)/\sqrt{P_i}$ is the transmit symbol of transmitter $i$, and the power constraint for each transmitter is normalized to unity; i.e., $E[|X_i(t)|^2]\leq 1$, $\forall i\in\{1,2,...,K\}$. The transmit power in the original channel model is absorbed in the channel coefficients, so that $\sqrt{P^{\alpha_{ki}}}$ and $\theta_{ki}$ are the magnitude and the phase, respectively, of the channel between transmitter $i$ and receiver $k$, $\forall i,k\in\{1,2,...,K\}$. We will call the exponent $\alpha_{ki}$ the channel strength level of the link between transmitter $i$ and receiver $k$. \textit{In the rest of the paper, we will only consider the equivalent channel model in (\ref{equ_channel}).}

Since this is a $K$-user interference channel, transmitter $i$ has message $W_i$ intended for receiver $i$, and  the messages $W_i$ are independent, $\forall i\in\{1,2,...,K\}$. We denote the size of the message set of user $i$ by $|W_i|$. For codewords spanning $n$ channel uses, the rates $R_i=\frac{\log|W_i|}{n}, i\in\{1,2,\cdots,K\}$, are achievable if the probability of error at all the receivers can be made arbitrarily small as $n$ approaches infinity. The channel capacity region $\mathcal{C}$ is the closure of the set of all achievable rate tuples. Collecting the channel strength levels and phases in the sets
\begin{equation}
\alpha\triangleq\{\alpha_{ki}\},~~~\theta\triangleq\{\theta_{ki}\},~~~\forall i,k\in\{1,2,...,K\},
\end{equation}
the capacity region is a function of $\alpha, \theta, P$, and is denoted as $\mathcal{C}(P,\alpha,\theta)$. 
\subsection{Generalized Degrees of Freedom}
The GDoF region of the $K$-user  interference channel as represented in (\ref{equ_channel}) is defined as
\begin{equation}
\begin{aligned}
\mathcal{D}(\alpha,\theta)\triangleq \Big\{(d_1,d_2,...,d_K): ~&d_i=\lim_{P\rightarrow\infty}\frac{R_i}{\log P},~~~\forall i\in\{1,2,...,K\},
&(R_1,R_2,...,R_K)\in \mathcal{C}(P,\alpha,\theta)\Big\}.
\end{aligned}
\end{equation}

In general, the channel capacity (GDoF) region of complex Gaussian interference channel may depend on both the channel strength levels $\alpha$, and the channel phases $\theta$. However, the capacity (GDoF) inner and outer bounds that we present in this paper depend \textit{only} on the channel strength levels $\alpha$. As such, our results hold regardless of whether or not the channel phase information is available to the transmitters.

\subsection{Capacity Region  within a Constant Gap}
 Following the same definition as in \cite{Tse_GDoF} and \cite{Yu_Cyclic}, an achievable region is said to be within $x$ bits of the capacity region if for any rate tuple $(R_1,R_2,...,R_K)$ on the boundary of the achievable region, the rate tuple $(R_1+x,R_2+x,...,R_K+x)$ is outside the channel capacity region. 
 
\subsection{Achievable Rate Region of TIN Scheme}
In the TIN scheme, transmitter $i$ uses a transmit power of $P^{r_i}$, $r_i\leq 0$ and each receiver treats all the incoming interference as noise, so that the SINR at receiver $i$ is given by
\begin{align*}
\text{SINR}_i=\frac{P^{\alpha_{ii}}\times P^{r_i}}{1+\sum_{j\neq i} P^{\alpha_{ij}} \times P^{r_j}}.
\end{align*}

This implies that the rate achieved by user $i$ through TIN is equal to
\begin{align}
R_i=\log(1+\text{SINR}_i)=\log\left(1+\frac{P^{\alpha_{ii}+r_i}}{1+\sum_{j\neq i} P^{\alpha_{ij}+r_j}}\right),
\end{align}
and therefore, the generalized degrees-of-freedom (GDoF) achieved by user $i$ equals
\begin{align}\label{eqq1}
d_i=\max\{0,\alpha_{ii}+r_i-\max\{0,\max_{j:j\neq i} (\alpha_{ij}+r_j)\}\}.
\end{align}

The achievable GDoF region through TIN, which we denote by $\mathcal{P}^*$, is the set of all $K$-tuples $(d_1,...,d_K)$ for which there exist $r_i$'s, $r_i\leq 0$, $i\in\{1,...,K\}$, such that (\ref{eqq1}) holds for all $i\in\{1,...,K\}$.

\section{Condition for Optimality of TIN}

The main result of this section is the following theorem, which introduces a  condition under which TIN is GDoF-optimal.

\begin{theorem}\label{th1}
In a $K$-user interference channel, where the channel strength level from transmitter $i$ to receiver $j$ is equal to $\alpha_{ji}$, $\forall i,j\in\{1,...,K\}$, if the following condition is satisfied
\begin{equation}\label{eqq10}
\alpha_{ii}\geq \max_{j:j\neq i}\{\alpha_{ji}\}+\max_{k:k\neq i}\{\alpha_{ik}\},~~~\forall i,j,k\in\{1,2,...K\},
\end{equation}
then power control and treating interference as noise can achieve the whole GDoF region. Moreover, the GDoF region is the set of all $K$-tuples $(d_1,d_2,...,d_K)$ satisfying
\begin{alignat}{2}
0\leq d_i&\leq \alpha_{ii},\:\:&&\forall i\in\{1,...,K\}\label{eqq11}\\
\sum_{j=1}^{m}d_{i_j}&\leq \sum_{j=1}^{m}(\alpha_{i_ji_j}-\alpha_{i_{j-1}i_j}),\:\:&&\forall (i_0,i_1,...,i_m)\in\Pi_K,~~\forall m\in\{2,3,...K\},\label{eqq12}
\end{alignat}
where $\Pi_K$ is the set of all possible cyclic sequences of all subsets of $\{1,...,K\}$, and the modulo-$m$ arithmetic is implicitly used on the user indices, e.g., $i_m=i_0$.
\end{theorem}

{\it Remark:} Condition (\ref{eqq10}) can be stated in words as --- \emph{for each user the desired signal strength is no less than the sum of  the strengths of the strongest interference from this user and the strongest interference to this user (all values in dB scale)}. Theorem \ref{th1} claims that under this condition, TIN is GDoF-optimal.

{\it Remark:} Both the condition (\ref{eqq10}) and the GDoF region specified by (\ref{eqq11})-(\ref{eqq12}) display a natural duality in the sense that they are both unchanged if the roles of the transmitters and receivers are switched, i.e., if all $\alpha_{ij}$ values are switched with $\alpha_{ji}$ values. In other words, for the same channel strengths, if we consider the reciprocal network (in the same sense as a multiple access channel being the reciprocal of a broadcast channel), then again under condition (\ref{eqq10}), TIN is GDoF-optimal, and the GDoF region is the same as in the original network. Such a duality holds also for the entire TIN region $\mathcal{P}^*$, and a similar duality relationship for the symmetric rate has been observed in \cite{Zander_Frodigh}.

\begin{ex}\label{ex1}
To  interpret the results in Theorem \ref{th1},  we derive and plot the GDoF region for a 3-user network in which the condition (\ref{eqq10}) is satisfied. Consider the 3-user network in Fig. \ref{TIN_ex1}.
\begin{figure}[h]
\centering
\subfigure[]{
\includegraphics[width= 4 cm]{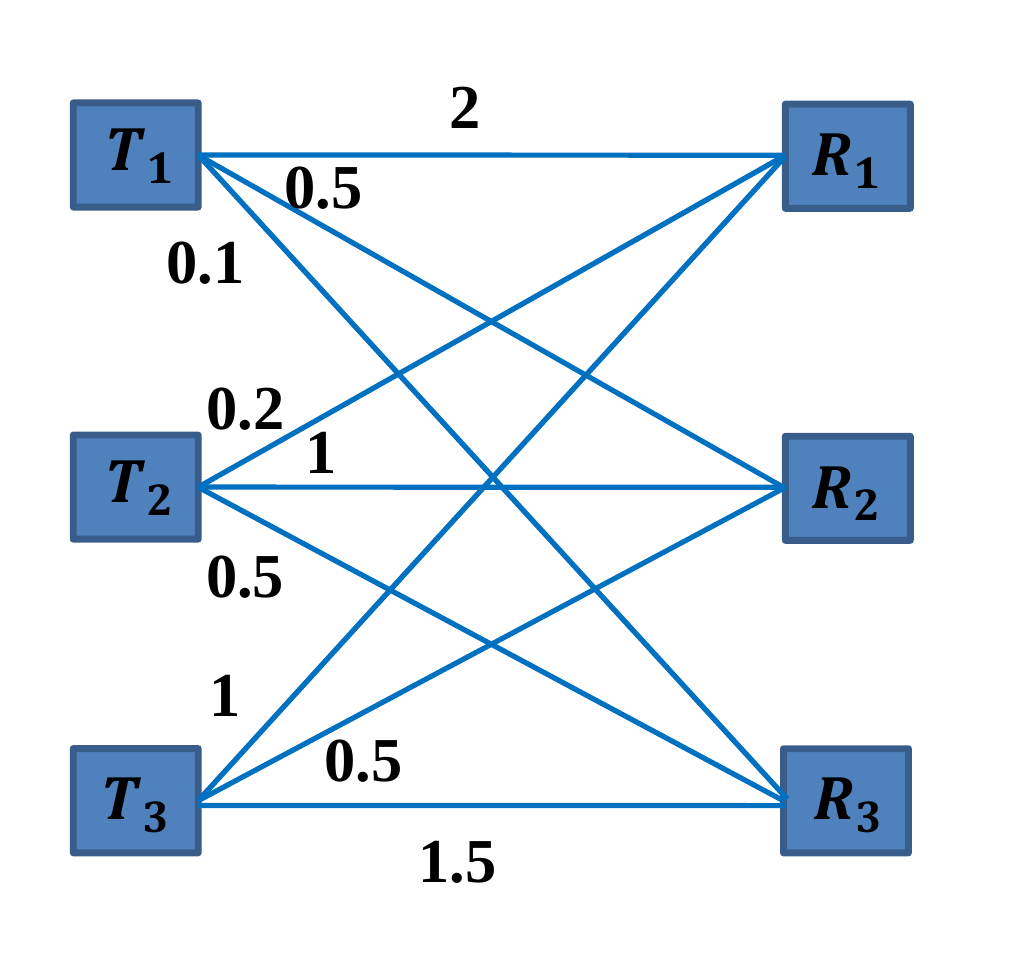}
\label{TIN_ex1}}
\hspace{1in}
\subfigure[]{
\includegraphics[trim=.61in 2.75in .8in 2.95in,clip,width=0.45\textwidth]{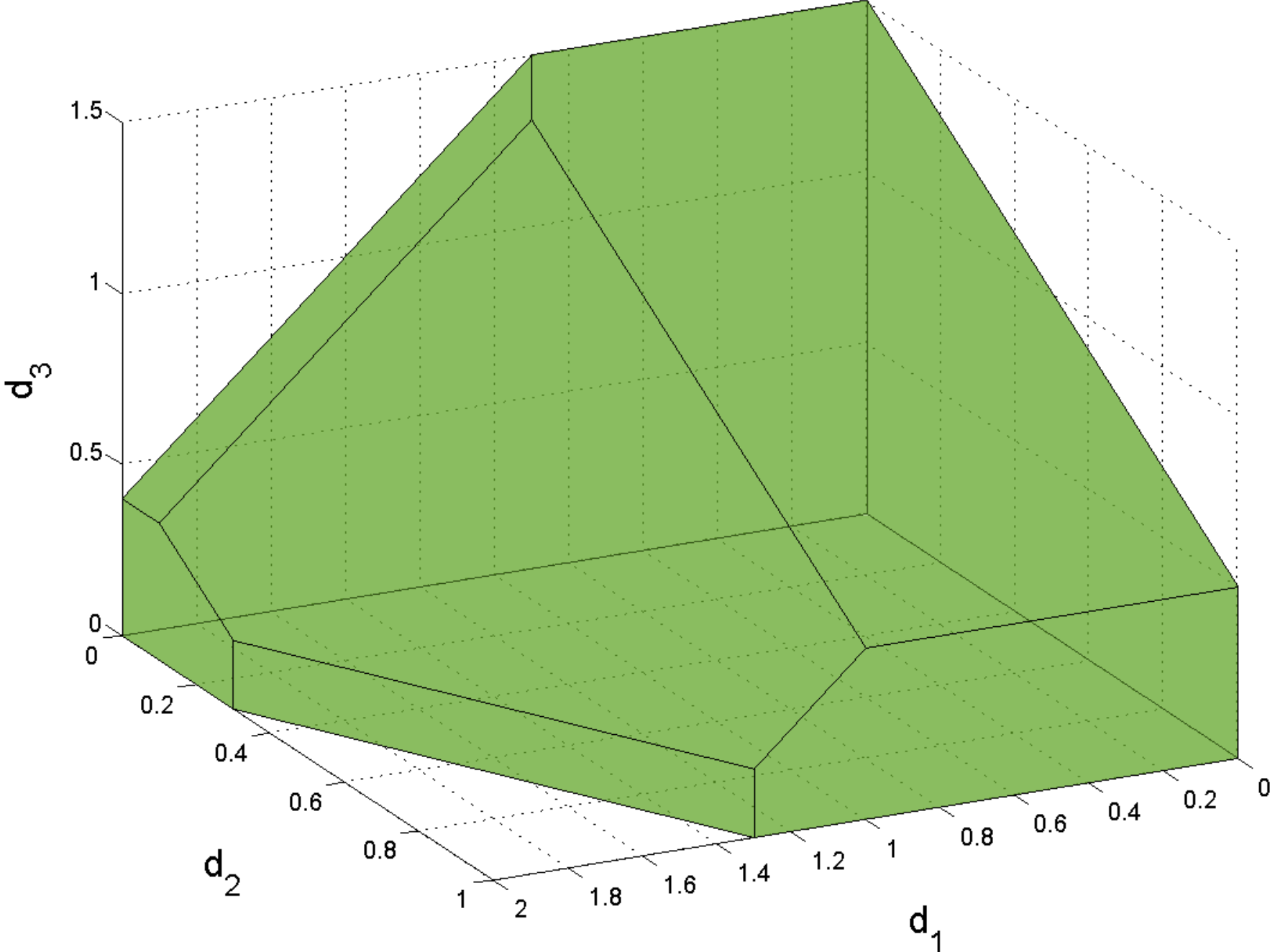}
\label{fig3}}
\caption[]{
\subref{TIN_ex1} A $3$-user interference channel, where the value on each link is equal to its channel strength level, and \subref{fig3} The GDoF region of this network, which is a convex polyhedron and can be achieved by TIN.}
\end{figure}
In this network, the channel strength level between transmitter $i$ and receiver $j$, $\alpha_{ji}$, is shown on the corresponding link, $\forall i,j\in\{1,2,3\}$. For the case of $K=3$, $\Pi_K=\{(1,2),(1,3),(2,3),\linebreak(1,2,3),(1,3,2)\}$. According to Theorem \ref{th1}, the GDoF region is the set of all $(d_1,d_2,d_3)$ satisfying
\begin{align*}
0\leq d_1&\leq 2\\
0\leq d_2&\leq 1\\
0\leq d_3&\leq 1.5\\
d_1+d_2&\leq 2.3\\
d_1+d_3&\leq 2.4\\
d_2+d_3&\leq 1.5\\
d_1+d_2+d_3&\leq 3.7\\
d_1+d_2+d_3&\leq 2.5,
\end{align*}
which is depicted in Fig. \ref{fig3}. Recall that the condition (\ref{eqq10}) is satisfied in the network of Fig. \ref{TIN_ex1} for all users $i\in\{1,2,3\}$. Therefore, Theorem \ref{th1} implies that TIN  achieves the entire GDoF region of this network. 

\end{ex}


We prove Theorem \ref{th1} through the following steps. We first show that under the condition stated in (\ref{eqq10}),  the achievable GDoF region of TIN simplifies into a polyhedral region. We study the polyhedral TIN region in some detail to understand its structure. In particular, we show that the polyhedral TIN region can be characterized by checking the existence of a potential function for an induced fully-connected directed graph, with nodes representing the source-destination pairs in the original interference channel (with the addition of a ``ground'' node) and a specific assignment of lengths to the arcs of the graph. Afterwards, we derive a dual characterization of the polyhedral TIN region and use the outer bounds developed in~\cite{Yu_Cyclic} to prove the optimality of polyhedral TIN, hence TIN, whenever condition (\ref{eqq10}) holds.

\subsection{Polyhedral Relaxation of TIN}
In the first step toward proving Theorem \ref{th1}, we introduce a polyhedral version of the TIN scheme. Ignoring the first $\max\{0,...\}$ term in (\ref{eqq1}) changes the scheme to a relaxed version, which we call the $\emph{polyhedral TIN}$ scheme. With this modification, the GDoF achieved by user $i$ will be
\begin{align}\label{eqq2}
d_i=\alpha_{ii}+r_i-\max\{0,\max_{j:j\neq i} (\alpha_{ij}+r_j)\},
\end{align}
and we denote the achievable GDoF region via polyhedral TIN by $\mathcal{P}$.

In general, comparing (\ref{eqq1}) and (\ref{eqq2}) shows that this modification can only shrink the achievable GDoF region of TIN. However, as we will show in the following, under the condition (\ref{eqq10}), the above relaxation incurs no loss in the GDoF region of TIN. In other words, \emph{when the condition (\ref{eqq10}) is satisfied, the TIN region $\mathcal{P}^*$ is equal to the polyhedral TIN region $\mathcal{P}$}. From (\ref{eqq2}), the polyhedral TIN region $\mathcal{P}$ can be characterized by a number of linear inequalities, which, as we will see, significantly contributes to understanding the TIN region $\mathcal{P}^*$. In fact, $\mathcal{P}$ is the set of all $K$-tuples $(d_1,...,d_K)$ for which there exist $r_i$'s, $i\in\{1,...,K\}$, such that
\begin{alignat}{2}
r_i&\leq 0,&&\forall i\in\{1,...,K\}\label{eqq3}\\
d_i\leq \alpha_{ii}+r_i \Leftrightarrow r_i&\geq d_i-\alpha_{ii},&&\forall i\in\{1,...,K\}\label{eqq4}\\
d_i\leq \alpha_{ii}+r_i-(\alpha_{ij}+r_j) \Leftrightarrow r_i-r_j&\geq \alpha_{ij}+(d_i-\alpha_{ii}),\:\:&&\forall i,j\in\{1,...,K\}, i\neq j.\label{eqq5}
\end{alignat}

As we will show, the region $\mathcal{P}$ can be fully characterized by (\ref{eqq11})-(\ref{eqq12}). Moreover,  as demonstrated in Example \ref{ex1}, the region $\mathcal{P}$  is a polyhedron, which is why the scheme is called polyhedral TIN. Note in general it is obvious that $\mathcal{P}\subseteq\mathcal{P}^*$, because TIN performs no worse than polyhedral TIN.

\subsection{Dual Characterization of Polyhedral TIN Region via Potential Functions}
Equipped with the aforementioned description of polyhedral TIN, we now characterize the polyhedral TIN region $\mathcal{P}$ for general channel strength levels. As mentioned earlier, $(d_1,d_2,...,d_K)\in\mathcal{P}$ if and only if there exist $r_i$'s, $i\in\{1,...,K\}$, satisfying
\begin{alignat}{2}
r_i&\leq 0,&&\forall i\in\{1,...,K\}\label{eqq13}\\
r_i&\geq d_i-\alpha_{ii},&&\forall i\in\{1,...,K\}\label{eqq14}\\
r_i-r_j&\geq \alpha_{ij}+(d_i-\alpha_{ii}),\:\:&&\forall i,j\in\{1,...,K\}, i\neq j.\label{eqq15}
\end{alignat}

\begin{figure}[h]
\centering
\subfigure[]{
\includegraphics[trim=3.2in 2.1in 3.2in 1.8in,clip,width=0.3\textwidth]{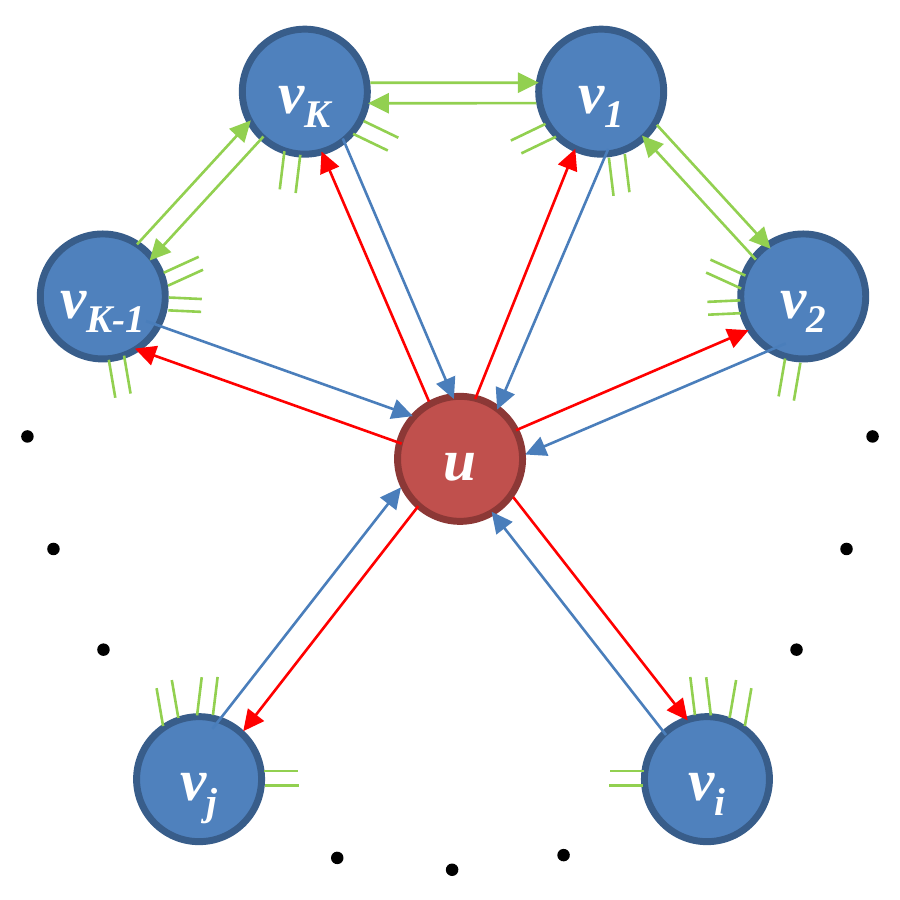}
\label{fig1}
}
\hspace{1in}
\subfigure[]{
\includegraphics[trim=3.2in 2.1in 3.3in 1.95in,clip,width=0.3\textwidth]{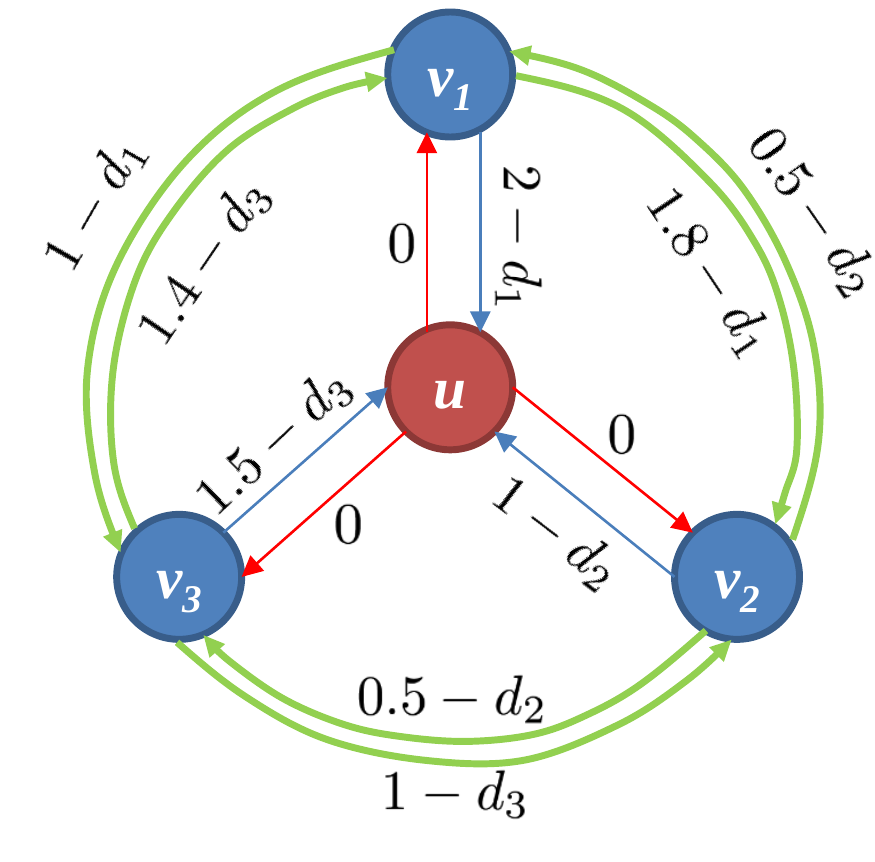}
\label{figD_ex}}
\caption[]{
\subref{fig1} The directed graph $D$ in which the green, blue and red arcs belong to $A_1$, $A_2$ and $A_3$, respectively. For simplicity, only some parts of the edges are shown in this figure. \subref{figD_ex} The corresponding directed graph $D$ for Example \ref{ex1}.}
\end{figure}

Now, we define a directed graph $D=(V,A)$, as shown in Fig. \ref{fig1}, where
\begin{align*}
V&=\{v_1,...,v_K,u\}\\
A&= A_1 \cup A_2 \cup A_3\\
A_1&=\{(v_i,v_j): i,j\in\{1,...,K\}, i\neq j\}\\
A_2&=\{(v_i,u): i\in\{1,...,K\}\}\\
A_3&=\{(u,v_i): i\in\{1,...,K\}\},
\end{align*}

\noindent and we assign a length $l(a)$ to every arc $a\in A$ as follows.
\begin{align*}
l(v_i,v_j)&=\alpha_{ii}-d_i-\alpha_{ij}\\
l(v_i,u)&=\alpha_{ii}-d_i\\
l(u,v_i)&=0.
\end{align*}

As an example, the corresponding directed graph $D$ for Example \ref{ex1} is drawn in Fig. \ref{figD_ex}. Evidently, this is a fully-connected directed graph, in which the length of each arc depends on the channel strength levels and the GDoFs we intend to achieve. This careful assignment of the lengths to the arcs of this graph allows us to use the following lemma.

\begin{lemma}\label{lem1}
If $\mathcal{P}$ denotes the polyhedral TIN region of a $K$-user interference channel, then\linebreak $(d_1,...,d_K)\in\mathcal{P}$ if and only if there exists a valid potential function for the graph $D$.
\end{lemma}

\textit{Proof of Lemma \ref{lem1}:} By definition \cite{Schrijver}, a function $p:V\rightarrow \mathbb{R}$ is called a potential if for every two nodes $a,b\in V$ such that $(a,b)\in A$, $l(a,b)\geq p(b)-p(a)$. These inequalities only depend on the \emph{difference} between potential function values. Therefore, without loss of generality, if there exists a valid potential function for the graph, we can make one node, say node $u$, $ground$; i.e., $p(u)=0$. Letting $r_i:=p(v_i)$, the potential function values should satisfy the following conditions.
\begin{align}
\alpha_{ii}-d_i-\alpha_{ij} &\geq r_j-r_i,\:\:&&\forall i,j\in\{1,...,K\}, i\neq j\\
\alpha_{ii}-d_i &\geq -r_i,\:\:&&\forall i\in\{1,...,K\}\\
0 &\geq r_i,\:\:&&\forall i\in\{1,...,K\}.
\end{align}

The above inequalities exactly match the ones in (\ref{eqq13})-(\ref{eqq15}). This completes the proof.
\hfill \QED

Next, we invoke the potential theorem of \cite{Schrijver}, re-stated below, to complete the characterization of the polyhedral TIN region, $\mathcal{P}$.

\bigskip
\textbf{Potential Theorem} [Theorem 8.2 of \cite{Schrijver}]\emph{: There exists a potential function for a directed graph $D$ if and only if each directed circuit in $D$ has nonnegative length.}
\bigskip

Combining Lemma \ref{lem1} and the potential theorem, we conclude that $(d_1,...,d_K)\in\mathcal{P}$ if and only if each directed circuit in the graph $D$ has nonnegative length. Therefore, it just remains to interpret the conditions of nonnegative length for the circuits.

We can categorize the circuits of $D$ in three classes:

\begin{itemize}
\item Circuits in the form of $(u,v_i,u)$. For these circuits, we have
\begin{align}\label{eqq16}
\alpha_{ii}-d_i\geq 0 \Leftrightarrow d_i\leq \alpha_{ii}.
\end{align}

\item Circuits in the form of $(v_{i_0},v_{i_1},...,v_{i_m})$, where ${i_0}={i_m}$, or in other words, the circuits which do not include node $u$. For these circuits, the nonnegative length condition will be
\begin{equation}\label{eqq17}
\begin{aligned}
\sum_{j=0}^{m-1} (\alpha_{i_j i_j}-d_{i_j}-\alpha_{i_j i_{j+1}})\geq 0 \Leftrightarrow&
\sum_{j=0}^{m-1} d_{i_j} \leq \sum_{j=0}^{m-1} (\alpha_{i_j i_j}-\alpha_{i_j i_{j+1}})\\
 \overset{(a)}{\Leftrightarrow}& \sum_{j=1}^{m}d_{i_j}\leq\sum_{j=1}^{m}(\alpha_{i_ji_j}-\alpha_{i_{j-1}i_j}).
\end{aligned}
\end{equation}
where in step $(a)$ we just reorder the terms in the right hand side and recall that $i_m=i_0$.

\item Circuits in the form of $(u,v_{i_1},...,v_{i_m},u)$, where $m>1$. For these circuits, the following inequality should hold.
\begin{align}\label{eqq18}
\sum_{j=1}^{m-1} (\alpha_{i_j i_j}-d_{i_j}-\alpha_{i_j i_{j+1}})+(\alpha_{i_m i_m}-d_{i_m})\geq 0.
\end{align}

Since $\alpha_{i_m i_1}\geq 0$, we have $\alpha_{i_m i_m}-d_{i_m} \geq \alpha_{i_m i_m}-d_{i_m}-\alpha_{i_m i_1}$. Therefore, given the conditions (\ref{eqq17}), the conditions in this class of circuits are redundant.
\end{itemize}

Consequently, we will end up with the conditions (\ref{eqq16})-(\ref{eqq17}), which coincide accurately with the conditions (\ref{eqq11})-(\ref{eqq12}), except for the non-negativity constraint on $d_i$'s, which is needed for the generalized degrees-of-freedom to be meaningful. This directly leads us to the following theorem which characterizes the polyhedral TIN region $\mathcal{P}$ for general channel strength levels.

\begin{theorem}\label{th2}
The GDoF region achieved through polyhedral TIN, denoted by $\mathcal{P}$, is the set of all $K$-tuples $(d_1,d_2,...,d_K)$ satisfying
\begin{alignat}{2}
0\leq d_i&\leq \alpha_{ii},\:\:&&\forall i\in\{1,...,K\}\label{eqq19}\\
\sum_{j=1}^{m}d_{i_j}&\leq \sum_{j=1}^{m}(\alpha_{i_ji_j}-\alpha_{i_{j-1}i_j}),\:\:&&\forall (i_0,i_1,...,i_m)\in\Pi_K,~~\forall m\in\{2,3,...K\},\label{eqq20}
\end{alignat}
where $\Pi_K$ is the set of all possible cyclic sequences of all subsets of $\{1,...,K\}$, and the modulo-$m$ arithmetic is implicitly used on the user indices, e.g., $i_m=i_0$.
\end{theorem}

Now, we are at a stage to complete the proof of Theorem \ref{th1}.

\subsection{Proof of Theorem \ref{th1}}

To prove Theorem \ref{th1}, we show that under the condition (\ref{eqq10}), the polyhedral TIN region $\mathcal{P}$ coincides with the GDoF region outer bound, therefore establishing the optimality of TIN under (\ref{eqq10}) and proving Theorem \ref{th1}. Note that from Theorem \ref{th2}, the region (\ref{eqq11})-(\ref{eqq12}) is exactly equal to the polyhedral TIN region $\mathcal{P}$ and therefore this GDoF region can be achieved by TIN. Therefore, we only need to prove the outer bounds on the GDoF region.

\begin{figure}[hbt]
\centering
\includegraphics[width= 3.5 cm]{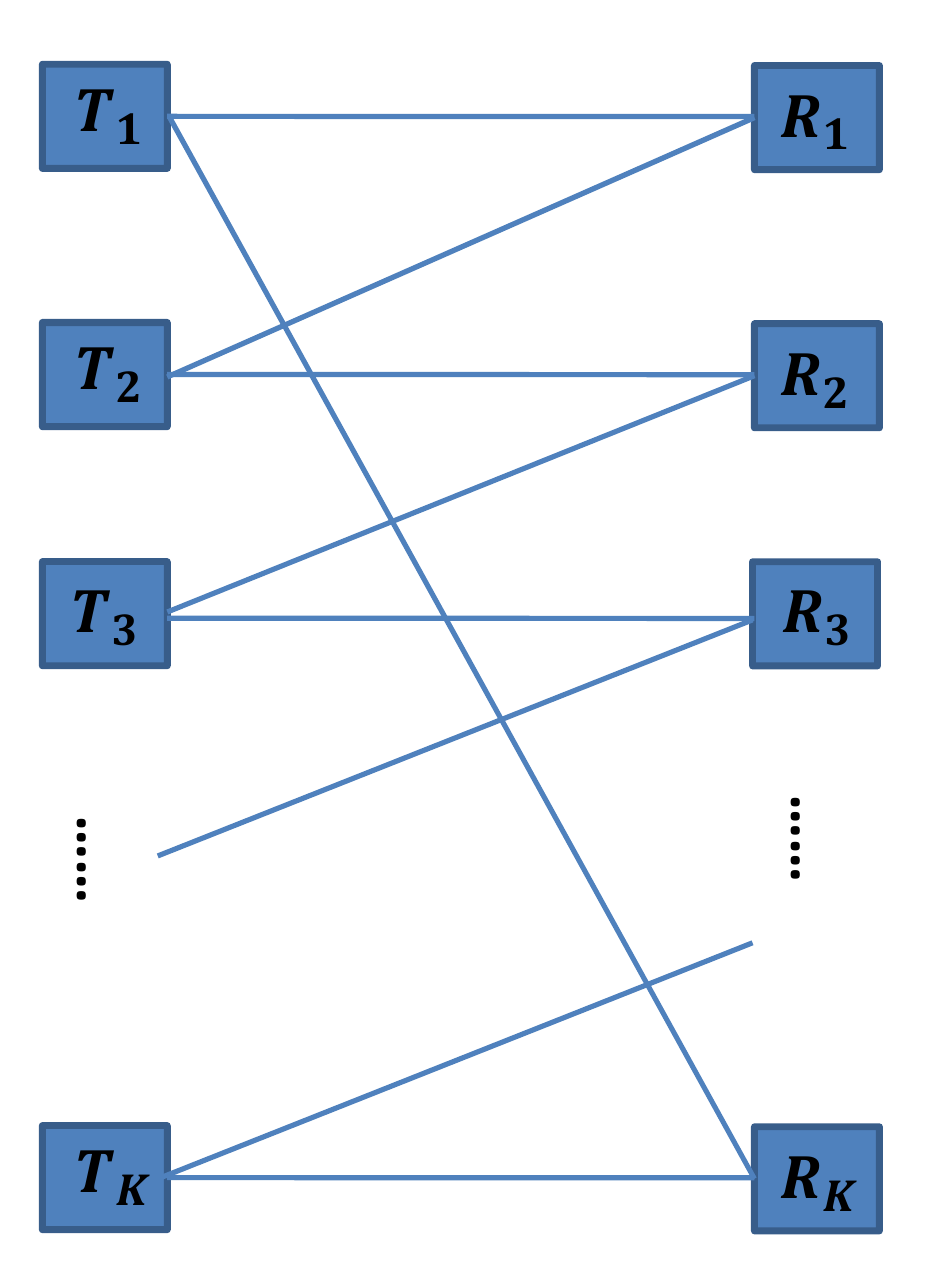}
\caption{$K$-user cyclic interference channel.}
\label{k_cyclic}
\end{figure}

In order to prove the converse, we use the outer bounds presented in \cite{Yu_Cyclic} for cyclic Gaussian interference channels. In \cite{Yu_Cyclic}, the authors investigate an interesting $K$-user cyclic Gaussian interference channel, where the $k$-th user only interferes with the $(k-1)$-th user (mod $K$) as shown in Fig. \ref{k_cyclic}. For completeness, let us re-state the key result of \cite{Yu_Cyclic} that we need to complete the proof.

For the $K$-user cyclic Gaussian interference channel, if we denote the channel gain between transmitter $i$ and receiver $j$ as $h_{j,i}$ and assume that each transmitter $i$ is subject to the power constraint $P_i$ and the additive white Gaussian noise (AWGN) at each receiver follows the distribution $\mathcal{CN}(0,\sigma^2)$, then we  define the signal-to-noise and interference-to-noise ratios for each user as follows
\begin{align*}
\mathrm{SNR_i}=\frac{|h_{i,i}|^2P_i}{\sigma^2},~~\mathrm{INR}_i=\frac{|h_{i-1,i}|^2P_i}{\sigma^2},~~\forall i\in\{1,2,...,K\},
\end{align*}
where modulo arithmetic is used on the user indices. The $K$-user cyclic Gaussian interference channel is in the \emph{weak} interference regime if
\begin{align*}
\mathrm{INR}_i\leq \mathrm{SNR}_i~~\forall i\in\{1,2,...,K\}.
\end{align*}

The following theorem gives the channel capacity outer bounds for this $K$-user cyclic channel in the weak interference regime.

  

\begin{theorem}\label{Yu_OB}
\emph{(Theorem 2 in \cite{Yu_Cyclic})}\footnote{There is a minor change of notation compared to \cite{Yu_Cyclic} in order to avoid confusion with the notations in this paper.} For the $K$-user cyclic Gaussian interference channel in the weak interference regime, the capacity region is included in the set of rate tuples $(R_1,R_2,...,R_K)$ such that
\begin{align}
R_i&\leq \lambda_i,\label{bound1}\\
\sum_{j=m}^{m+l-1}R_j&\leq \min\Big\{\gamma_m+\sum_{j=m+1}^{m+l-2}\kappa_j+\beta_{m+l-1},~\mu_m+\sum_{j=m}^{m+l-2}\kappa_j+\beta_{m+l-1}\Big\},\label{bound2}\\
\sum_{j=1}^KR_j&\leq\min\Big\{\sum_{j=1}^K\kappa_{j},\rho_1,\rho_2,...,\rho_K\Big\},\label{bound3}\\
\sum_{j=1}^KR_j+R_i&\leq\beta_i+\gamma_i+\sum_{j=1,j\neq i}^K\kappa_j,
\end{align}
where $i$,$m$ $\in\{1,2,...,K\}$, $l\in\{2,3,...,K-1\}$, and
\begin{align*}
\kappa_i&=\log\left(1+\mathrm{INR}_{i+1}+\frac{\mathrm{SNR}_i}{1+\mathrm{INR}_i}\right)\\
\beta_i&=\log\left(\frac{1+\mathrm{SNR}_i}{1+\mathrm{INR}_i}\right)\\
\gamma_i&=\log(1+\mathrm{INR}_{i+1}+\mathrm{SNR}_i)\\
\lambda_i&=\log(1+\mathrm{SNR}_i)\\
\mu_i&=\log(1+\mathrm{INR}_i)\\
\rho_i&=\beta_{i-1}+\gamma_i+\sum_{j=1,j\not\in\{i,i-1\}}^K\kappa_j.
\end{align*}
\end{theorem}

Equipped with Theorem \ref{Yu_OB}, we can now prove the converse for Theorem \ref{th1}. The individual bounds (\ref{eqq11}) follow directly from the inequalities (\ref{bound1}). In fact, from (\ref{bound1}) we have
\begin{align*}
d_i=\lim_{P\rightarrow\infty}\frac{R_i}{\log P} \leq \lim_{P\rightarrow\infty}\frac{\log(1+P^{\alpha_{ii}})}{\log P}=\alpha_{ii},
\end{align*}
for any $i\in\{1,2,...,K\}$. Also, the cyclic outer bounds (\ref{eqq12}) follow from the outer bounds (\ref{bound3}) under the condition (\ref{eqq10}). Under this condition, for any cycle $(i_0,i_1,...,i_m)\in\Pi_K$ we have
\begin{align*}
\lim_{P\rightarrow\infty}\frac{\sum_{j=1}^m \kappa_{i_j}}{\log P}
&=\lim_{P\rightarrow\infty}\frac{\sum_{j=1}^m \log\left(1+P^{\alpha_{i_j i_{j+1}}}+\frac{P^{\alpha_{i_j i_j}}}{1+P^{\alpha_{i_{j-1} i_j}}}\right)}{\log P}\\
&=\sum_{j=1}^m \max\{0,\alpha_{i_j i_{j+1}},\alpha_{i_ji_j}-\alpha_{i_{j-1}i_j}\}\\
&=\sum_{j=1}^m (\alpha_{i_ji_j}-\alpha_{i_{j-1}i_j}),\numberthis\label{kappa}
\end{align*}
and it follows that for any $k\in\{1,2,...,m\}$,
\begin{align*}
\lim_{P\rightarrow\infty}\frac{\rho_{i_k}}{\log P}
&=\lim_{P\rightarrow\infty}
\frac{\beta_{i_{k-1}}+\gamma_{i_k}+\sum_{j=1,j\not\in\{k,k-1\}}^m\kappa_{i_j}}{\log P}\\
&=\lim_{P\rightarrow\infty}
\frac{\log\left(\frac{1+P^{\alpha_{i_{k-1} i_{k-1}}}}{1+P^{\alpha_{i_{k-2} i_{k-1}}}}\right)+\log(1+P^{\alpha_{i_k i_{k+1}}}+{P^{\alpha_{i_k i_k}}})}{\log P}
+\sum_{j=1,j\not\in\{k,k-1\}}^m (\alpha_{i_ji_j}-\alpha_{i_{j-1}i_j})\\
&=(\alpha_{i_{k-1} i_{k-1}}-\alpha_{i_{k-2} i_{k-1}})+
\alpha_{i_k i_k}+\sum_{j=1,j\not\in\{k,k-1\}}^m (\alpha_{i_ji_j}-\alpha_{i_{j-1}i_j})\\
&=\alpha_{i_k i_k}+\sum_{j=1,j\not\in\{k\}}^m (\alpha_{i_ji_j}-\alpha_{i_{j-1}i_j}).\numberthis\label{rho}
\end{align*}

Therefore, comparing (\ref{kappa}) and (\ref{rho}) implies that under the condition (\ref{eqq10}), we have
\begin{align*}
\lim_{P\rightarrow\infty}\frac{\sum_{j=1}^m \kappa_{i_j}}{\log P}\leq \lim_{P\rightarrow\infty}\frac{\rho_{i_k}}{\log P},~~~\forall k\in\{1,2,...,m\},
\end{align*}
and then the outer bound (\ref{bound3}) implies that
\begin{align*}
\sum_{j=1}^{m}d_{i_j}&\leq \lim_{P\rightarrow\infty}\frac{\sum_{j=1}^m \kappa_{i_j}}{\log P}=\sum_{j=1}^m (\alpha_{i_ji_j}-\alpha_{i_{j-1}i_j}),
\end{align*}
for any cycle $(i_0,i_1,...,i_m)\in\Pi_K$. This completes the outer bound. \hfill \QED

\bigskip

Note that as we explained before, when the condition (\ref{eqq10}) is satisfied, the TIN region $\mathcal{P}^*$ is equal to the polyhedral TIN region $\mathcal{P}$, which is a convex polyhedron as shown in Theorem \ref{th2}. This means that in this regime, time-sharing \emph{cannot} help enlarge the GDoF achievable region via TIN.

\section{Constant Gap to Capacity}
In this section, we show that when  condition (\ref{eqq10}) holds, so that TIN is GDoF-optimal, we can apply the insight gained in the GDoF study to prove that TIN can also achieve the whole channel capacity region to within a constant gap at any finite SNR. The main result of this section is mentioned in the following theorem.

\begin{theorem}\label{constant_gap}
In a $K$-user interference channel, where the channel strength level between transmitter $i$ and receiver $j$ is $\alpha_{ji}$, if condition (\ref{eqq10}) holds,
then TIN can achieve to within $\log_2(3K)$ bits of the capacity region.
\end{theorem}

\textit{Proof}: (\textbf{Converse}) Using Theorem \ref{Yu_OB}, we obtain the following outer bounds.
\begin{alignat}{2}
R_i&\leq\log_2(1+P^{\alpha_{ii}}),&&\forall i\in\{1,2,...,K\}\\
\sum_{j=1}^mR_{i_j}\leq\sum_{j=1}^m\log_2(&1+P^{\alpha_{i_ji_{j+1}}}+\frac{P^{\alpha_{i_ji_j}}}{1+P^{\alpha_{i_{j-1}i_j}}}),~~&&\forall (i_0,i_1,...,i_m)\in\Pi_K,~~\forall m\in\{2,3,...K\}.
\end{alignat}

Since $P>1$, it follows that
\begin{equation}
\label{Rate_OB1}
R_i\leq\log_2(1+P^{\alpha_{ii}})\leq\alpha_{ii}\log_2P+1,~~~~\forall i\in\{1,2,...,K\}
\end{equation}
\begin{equation}
\label{Rate_OB2}
\begin{aligned}
\sum_{j=1}^mR_{i_j}&\leq\sum_{j=1}^m\log_2(1+P^{\alpha_{i_ji_{j+1}}}+\frac{P^{\alpha_{i_ji_j}}}{1+P^{\alpha_{i_{j-1}i_j}}})\\
&<\sum_{j=1}^m\log_2(1+P^{\alpha_{i_ji_{j+1}}}+\frac{P^{\alpha_{i_ji_j}}}{P^{\alpha_{i_{j-1}i_j}}})\\
&=\sum_{j=1}^m\log_2(\frac{P^{\alpha_{i_{j-1}i_j}}+P^{\alpha_{i_ji_{j+1}}+\alpha_{i_{j-1}i_j}}+P^{\alpha_{i_ji_j}}}{P^{\alpha_{i_{j-1}i_j}}})\\
&\leq\sum_{j=1}^m\log_2(\frac{3P^{\alpha_{i_ji_j}}}{P^{\alpha_{i_{j-1}i_j}}})\\
&=\sum_{j=1}^m[(\alpha_{i_ji_j}-\alpha_{i_{j-1}i_j})\log_2P+\log_23],
\end{aligned}
\end{equation}
for all cycles $(i_0,i_1,...,i_m)\in\Pi_K$, $\forall m\in\{2,3,...K\}$.

(\textbf{Achievability}) Consider the power control and TIN scheme, where the power allocated to each transmitter is equal to $P^{r_i}$ ($r_i\leq 0$, $\forall i\in\{1,2,...,K\}$), and the achievable rate for each user is
\begin{equation}
R_{i,\mathrm{TIN}}=\log_2(1+\frac{P^{r_i+\alpha_{ii}}}{1+\sum_{j\neq i}P^{r_j+\alpha_{ij}}}).
\end{equation}

From the proof of Theorem \ref{th1}, we know that under the condition (\ref{eqq10}), if $d_i$'s satisfy (\ref{eqq11}) and (\ref{eqq12}), then there exist $r_i$'s such that
\begin{alignat}{2}
r_i+\alpha_{ii}-\max_{j\neq i}\{0,r_{j}+\alpha_{ij}\}&= d_i,~~~&&\forall i,j\in\{1,2,...,K\}\label{con_cap_1}\\
r_i&\leq 0,~~~&&\forall i\in\{1,2,...,K\}.\label{con_cap_2}
\end{alignat}

Therefore, we can write
\begin{equation}
\label{R_TIN}
\begin{aligned}
R_{i,\mathrm{TIN}}=&\log_2(1+\frac{P^{r_i+\alpha_{ii}}}{1+\sum_{j\neq i}P^{r_j+\alpha_{ij}}})\\
\geq&\log_2(\frac{P^{r_i+\alpha_{ii}}}{P^0+\sum_{j\neq i}P^{r_j+\alpha_{ij}}})\\
\geq&\log_2(\frac{P^{r_i+\alpha_{ii}}}{KP^{r_i+\alpha_{ii}-d_i}})\\
=&d_i\log_2P+\log_2(\frac{1}{K}).
\end{aligned}
\end{equation}

In other words, when $d_i$'s satisfy (\ref{eqq11}) and (\ref{eqq12}), the rates in (\ref{R_TIN}) are always achievable by TIN, $\forall i\in\{1,...,K\}$. Therefore, the achievable rate region by TIN includes the rate tuples $(R_{1,\mathrm{TIN}},R_{2,\mathrm{TIN}},...,R_{K,\mathrm{TIN}})$ satisfying
\begin{align}
R_{i,\mathrm{TIN}}&\leq\alpha_{ii}\log_2P+\log_2(\frac{1}{K})\label{Rate_IB1}~~~~\forall i\in\{1,2,...,K\}\\
\sum_{j=1}^mR_{i_j,\mathrm{TIN}}&=\sum_{j=1}^m[d_{i_j}\log_2P+\log_2(\frac{1}{K})]\nonumber\\
&\leq\sum_{j=1}^m[(\alpha_{i_ji_j}-\alpha_{i_{j-1}i_j})\log_2P+\log_2(\frac{1}{K})],\label{Rate_IB2}
\end{align}
for all cycles $(i_0,i_1,...,i_m)\in\Pi_K$, $\forall m\in\{2,3,...K\}$.

Comparing (\ref{Rate_OB1})-(\ref{Rate_OB2}) with (\ref{Rate_IB1})-(\ref{Rate_IB2}), we can characterize the approximate channel capacity to within a constant gap, which is only dependent on the number of users $K$. We can show that TIN achieves to within $\log_2(3K)$ bits of the capacity region. To this end, we need to show that each of the rate constraints in (\ref{Rate_IB1}) and (\ref{Rate_IB2}) is within $\log_2(3K)$ bits of its corresponding outer bound in (\ref{Rate_OB1}) and (\ref{Rate_OB2}), i.e., the following inequalities always hold\footnote{Notice that since in the second line of (\ref{Rate_OB2}) there exists a ``$<$'', ``$\leq$'' is fine for the second inequality in (\ref{gap}).},
\begin{equation}
\label{gap}
\begin{aligned}
\sigma_{R_i}&< \log_2(3K),~~~~~~\forall i\in\{1,2,...,K\}\\
\sigma_{\sum_{j=1}^mR_{i_j}}&\leq m\log_2(3K),~~~\forall(i_0,i_1,...,i_m)\in\Pi_K,~\forall m\in\{2,3,...,K\},
\end{aligned}
\end{equation}
where $\sigma_{(.)}$ denotes the difference between the achievable rate in (\ref{Rate_IB1}) and (\ref{Rate_IB2}) and its corresponding outer bound in (\ref{Rate_OB1}) and (\ref{Rate_OB2}). For $\sigma_{R_i}$, we have
\begin{equation}
\label{inequ1}
\begin{aligned}
\sigma_{R_i}&=[\alpha_{ii}\log_2P+1]-[\alpha_{ii}\log_2P+\log_2(\frac{1}{K})]\\
&=1+\log_2K<\log_2(3K),
\end{aligned}
\end{equation}
and for $\sigma_{\sum_{j=1}^mR_{i_j}}$, we have,
\begin{equation}
\label{inequ2}
\begin{aligned}
\sigma_{\sum_{j=1}^mR_{i_j}}&=\sum_{j=1}^m[(\alpha_{i_ji_j}-\alpha_{i_{j-1}i_j})\log_2P+\log_23]-\sum_{j=1}^m[(\alpha_{i_ji_j}-\alpha_{i_{j-1}i_j})\log_2P+\log_2(\frac{1}{K})]\\
&=\sum_{j=1}^m[\log_23+\log_2K]=m\log_2(3K).
\end{aligned}
\end{equation}

Since (\ref{inequ1}) and (\ref{inequ2}) hold for all ranges of $i$ and $m$, the proof is complete. \hfill \QED

\section{The General Achievable GDoF Region of TIN}
In this section, we remove the constraint (\ref{eqq10}) on the channel gains, and investigate the achievable GDoF region by TIN for $K$-user interference channels with general channel strength levels. As we show, the TIN region $\mathcal{P}^*$ is equal to the union of multiple polyhedra, each of which is in the form of the polyhedral TIN region of a subset of the users of the network. Remarkably, the TIN region is almost the same as the polyhedral TIN region in the sense that the measure of the difference of the two sets is zero in $\mathbb{R}^K$.

We have shown that when (\ref{eqq10}) holds, the original TIN region $\mathcal{P}^*$ is equal to the polyhedral TIN region $\mathcal{P}$. Now, the natural question to ask is what the TIN region $\mathcal{P}^*$ is for $K$-user interference channels with general channel strength levels. The following theorem settles this issue.

\begin{theorem}\label{General_TIN_Region}
In a $K$-user interference channel, where the channel strength level from transmitter $i$ to receiver $j$ is equal to $\alpha_{ji}$, the achievable GDoF region through power control and treating interference as noise, denoted by $\mathcal{P}^*$, is equal to
\begin{align}\label{true_tin_region}
\mathcal{P}^*=\bigcup_{\mathcal{S}\subseteq \{1,...,K\}} \mathcal{P}_\mathcal{S},
\end{align}
where $\mathcal{P}_\mathcal{S}$, $\mathcal{S}\subseteq \{1,...,K\}$, is defined as
\begin{align*}
\mathcal{P}_\mathcal{S}=\{(d_1,...,d_K)&: d_i=0, \forall i\in\mathcal{S}, 0\leq d_j\leq \alpha_{jj},\forall j\in\mathcal{S}^c, \\
&\qquad\sum_{j=1}^m d_{i_j}\leq \sum_{j=1}^m (\alpha_{i_j i_j}-\alpha_{i_{j-1} i_{j}}),\forall (i_0,i_1,i_2,...,i_m)\in\Pi_{\mathcal{S}^c}\},
\end{align*}
and $\Pi_{\mathcal{S}^c}$ is the set of all possible cyclic sequences of all subsets of $\mathcal{S}^c$.
\end{theorem}

In words, the TIN region $\mathcal{P}^*$ is the union of the polyhedral TIN regions $\mathcal{P}_\mathcal{S}$, each of which corresponds to the case where the users in $\mathcal{S}$ are made silent. The proof is given in the appendix.

As Theorem \ref{General_TIN_Region} shows, the TIN region $\mathcal{P}^*$ is almost the same as the polyhedral TIN region $\mathcal{P}$ in the sense that the measure of the difference of the two sets is zero in $\mathbb{R}^K$. Furthermore, as opposed to the polyhedral TIN region, the TIN region may not be convex in general, and if time-sharing is allowed alongside with TIN, the achievable region may become substantially larger. Therefore, the above theorem also reveals that \emph{when the sufficient condition (\ref{eqq10}) is violated, time-sharing may help enlarge the achievable GDoF region of TIN.}

\begin{ex}
Consider the $3$-user cyclic channel shown in Fig. \ref{TIN_ex2}. Notice that for user 3 the sufficient condition (\ref{eqq10}) does not hold.

\begin{figure}[h]
\begin{center}
 \includegraphics[width= 4.8 cm]{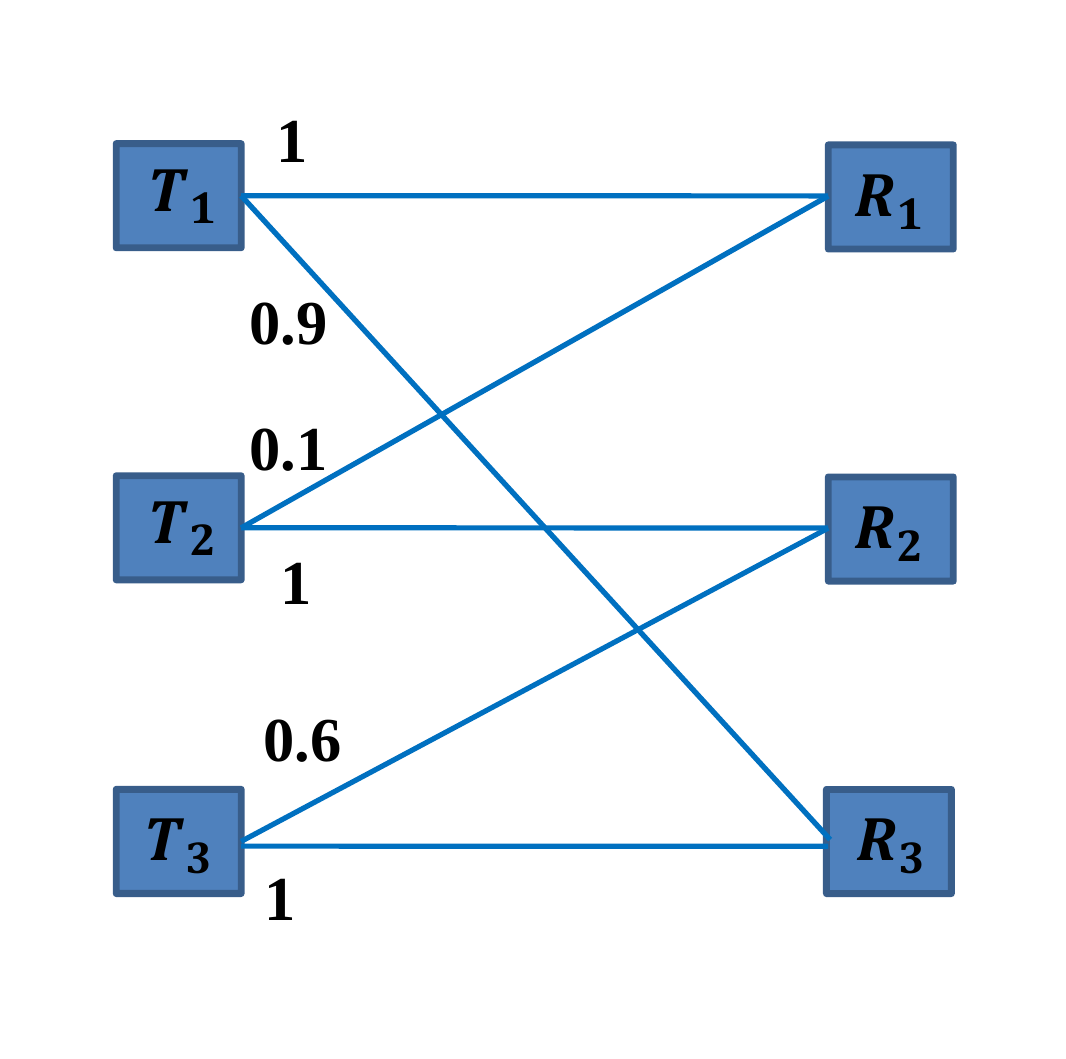}
 \caption{A 3-user cyclic channel, where the strength levels for each link is shown in the figure.}
\label{TIN_ex2}
\end{center}
\end{figure}

First, if all the users are active, we can get the polyhedral TIN region as follows.
\begin{equation}
\begin{aligned}
\mathcal{P}_{\emptyset}=\big\{(d_1,d_2,d_3):~& 0\leq d_i\leq1,~\forall i\in\{1,2,3\},\\
&d_1+d_2\leq1.9,~d_2+d_3\leq1.4,~d_1+d_3\leq1.1,\\
&d_1+d_2+d_3\leq1.4\big\},
\end{aligned}
\end{equation}
which is in fact the polyhedral TIN region $\mathcal{P}$ we defined earlier.

Then, consider the cases in which only one of the three users is made silent and hence has GDoF zero, and the other two users are active. In such cases, we only need to consider the Z-channel between the remaining two users, implying that
\begin{align*}
\mathcal{P}_{\{1\}}&=\big\{(d_1,d_2,d_3):~ d_1=0,~0\leq d_2\leq1,~0\leq d_3\leq1,~d_2+d_3\leq1.4\big\}\\
\mathcal{P}_{\{2\}}&=\big\{(d_1,d_2,d_3):~ d_2=0,~0\leq d_1\leq1,~0\leq d_3\leq1,~d_1+d_3\leq1.1\big\}\\
\mathcal{P}_{\{3\}}&=\big\{(d_1,d_2,d_3):~ d_3=0,~0\leq d_1\leq1,~0\leq d_2\leq1,~d_1+d_2\leq1.9\big\}.
\end{align*}

It is easy to verify that
\begin{equation*}
\mathcal{P}_{\{1\}}\subseteq \mathcal{P}_{\emptyset},~\mathcal{P}_{\{2\}}\subseteq \mathcal{P}_{\emptyset},
\end{equation*}
but
\begin{equation*}
\mathcal{P}_{\{3\}}\not\subseteq \mathcal{P}_{\emptyset}.
\end{equation*}

For instance, the GDoF tuple $(1,0.9,0)\in \mathcal{P}_{\{3\}}$ is not in the GDoF region $\mathcal{P}_{\emptyset}$ since it violates the cycle bound $d_1+d_2+d_3\leq 1.4$.

Next, consider the cases in which two users are made silent.
\begin{align*}
\mathcal{P}_{\{2,3\}}&=\big\{(d_1,d_2,d_3):~0\leq d_1\leq 1,~d_2=d_3=0\big\}\\
\mathcal{P}_{\{1,3\}}&=\big\{(d_1,d_2,d_3):~0\leq d_2\leq 1,~d_1=d_3=0\big\}\\
\mathcal{P}_{\{1,2\}}&=\big\{(d_1,d_2,d_3):~0\leq d_3\leq 1,~d_1=d_2=0\big\},
\end{align*}
and it can be verified that
\begin{equation*}
\mathcal{P}_{\{2,3\}}\subseteq \mathcal{P}_{\emptyset},~\mathcal{P}_{\{1,3\}}\subseteq \mathcal{P}_{\emptyset},~\mathcal{P}_{\{1,2\}}\subseteq \mathcal{P}_{\emptyset}.
\end{equation*}

Finally, we have
\begin{equation*}
\mathcal{P}_{\{1,2,3\}}=\big\{(d_1,d_2,d_3):~d_1=d_2=d_3=0\big\}\subseteq \mathcal{P}_{\emptyset}.
\end{equation*}

Therefore, the TIN region is equal to
\begin{equation}
\mathcal{P}^*=\mathcal{P}_{\emptyset}\cup \mathcal{P}_{\{1\}}\cup \mathcal{P}_{\{2\}}\cup \mathcal{P}_{\{3\}}\cup \mathcal{P}_{\{1,2\}}\cup \mathcal{P}_{\{2,3\}}\cup \mathcal{P}_{\{1,3\}} \cup \mathcal{P}_{\{1,2,3\}}=\mathcal{P}_{\emptyset}\cup \mathcal{P}_{\{3\}}.
\end{equation}

This region is illustrated in Fig. \ref{true_tin}, where the yellow region corresponds to $\mathcal{P}_{\emptyset}$ and the blue region corresponds to $\mathcal{P}_{\{3\}}$. Note that since for user $3$, the sufficient condition (\ref{eqq10}) is violated, the polyhedral TIN region $\mathcal{P}=\mathcal{P}_{\emptyset}$ is not the whole GDoF region for this 3-user cyclic channel. Moreover, as Fig. \ref{true_tin} shows, the region $\mathcal{P}^*$ is not convex. Therefore, time-sharing between $\mathcal{P}_{\emptyset}$ and $\mathcal{P}_{\{3\}}$ can help enlarge the achievable GDoF region via TIN.
\begin{figure}[h]
\centering
\includegraphics[trim=.61in 2.6in .8in 2.95in,clip,width=0.53\textwidth]{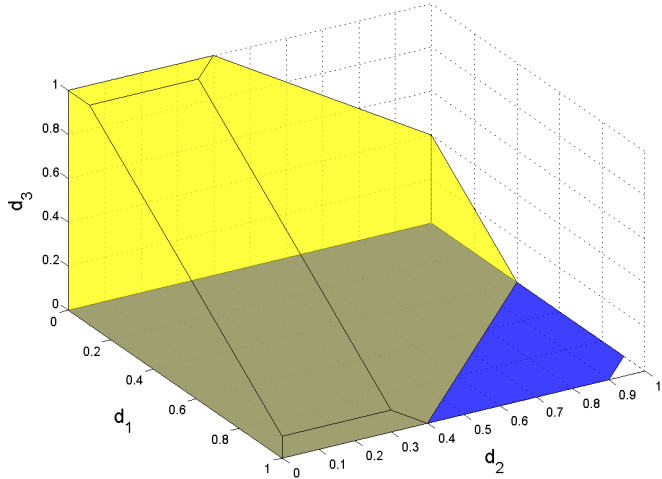}
\caption{The TIN region of the network in Figure \ref{TIN_ex2}, which is the union of the yellow region ($\mathcal{P}_{\emptyset}$) and the blue region ($\mathcal{P}_{\{3\}}$).}
\label{true_tin}
\end{figure}
\end{ex}

\section{Numerical Analysis}
\begin{figure}[hbt]
\centering
\subfigure[]{
\includegraphics[trim=1.65in 5.5in 1.7in 0.4in,clip,width=0.35\textwidth]{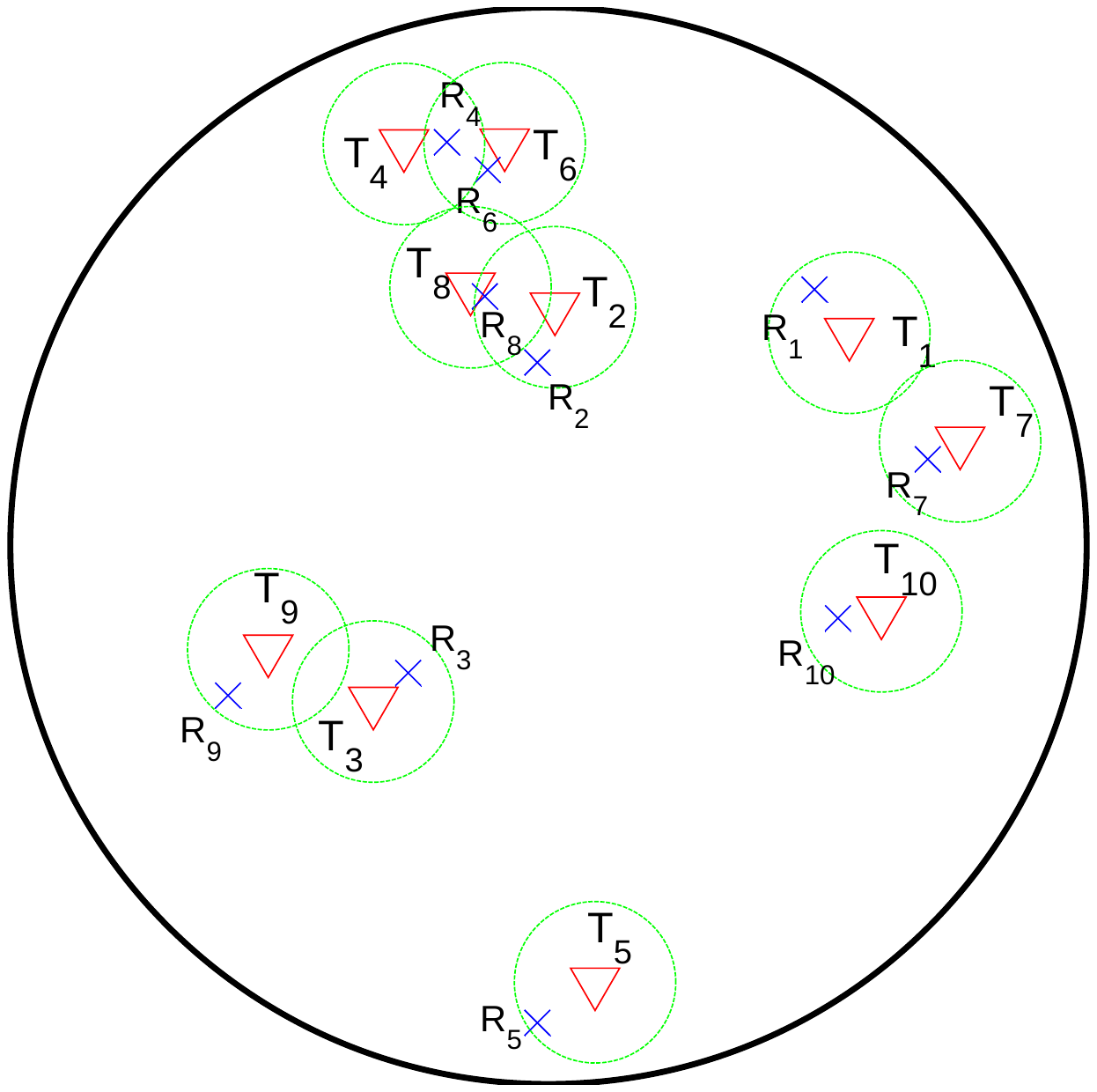}
\label{ex_net_sim}}~~~~~~~~~~~~
\subfigure[]{
\includegraphics[trim=0.6in 2.7in 0.7in 2.8in,clip,width=0.465\textwidth]{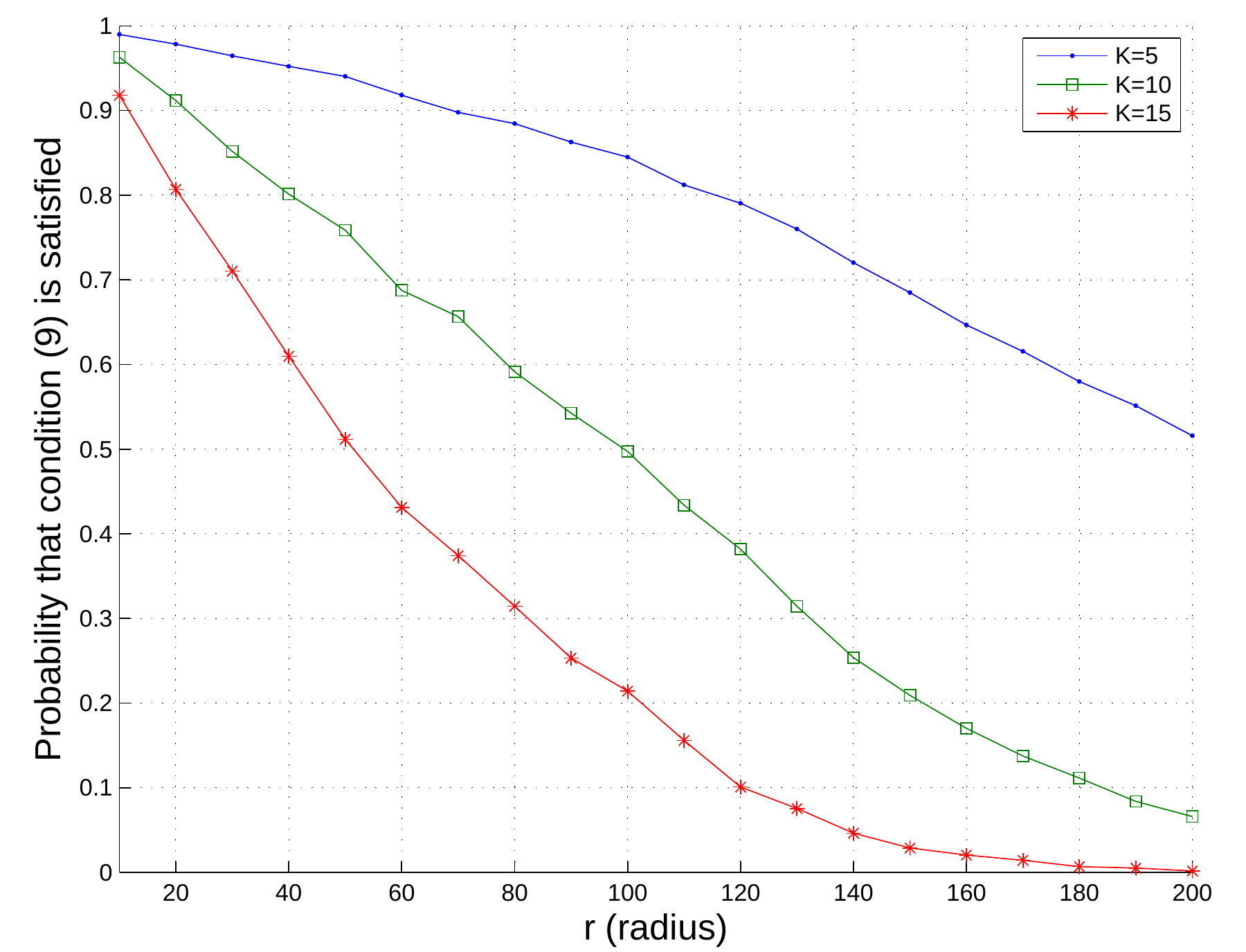}
\label{sim_result}}
\caption[]{
\subref{ex_net_sim} A 10-user interference channel where the black circle, green circles, red triangles, and blue crosses represent the whole cell area, the coverage area of base stations, base stations (transmitters), and receivers, respectively. The coverage radius of each transmitter is taken to be $r=100$m. \subref{sim_result} Effect of the coverage radius and the number of users on the probability that the sufficient condition (\ref{eqq10}) is satisfied.}
\end{figure}

In this section, we numerically compute the probability that the sufficient condition (\ref{eqq10}) is satisfied in a typical wireless scenario. We consider a circular cell with a radius of 1 km and place $K$ base stations (transmitters) randomly and uniformly over the cell area. Each base station is assumed to have a coverage radius of $r$. In order to create a $K$-user interference channel with strong enough direct links, we consider $K$ mobile receivers such that the $i$-th mobile receiver is located randomly and uniformly inside the coverage area of the $i$-th base station, $i\in\{1,2,...,K\}$. A realization of such a network scenario is depicted in Fig. \ref{ex_net_sim}.

For the channel gain values, we make use of the Erceg model \cite{erceg}, operating at a frequency of 2GHz and using the terrain category of hilly/light tree density. Taking the noise floor as -110 dBm, we choose the transmit power of all the base stations such that the expected value of the SNR at the boundary of their coverage area is 0 dB. Then, we randomly locate the base stations and mobile receivers according to the coverage radius $r$. Fig. \ref{sim_result} demonstrates the result of our numerical analysis.

As illustrated in this plot, the probability that the sufficient condition (\ref{eqq10}) for the GDoF-optimality of TIN is satisfied decreases as the density of the network increases, either by increasing the number of users or by increasing the coverage radius of each base station. However, as a typical scenario, it is noteworthy that for the case of a 10-user interference channel with the coverage radius of 100m for each base station, the sufficient condition (\ref{eqq10}) is satisfied half the times. This means that with a probability of 50\%, TIN is GDoF-optimal and can also achieve the whole capacity region of the network to within a constant gap. It therefore implies that the sufficient condition (\ref{eqq10}) can be actually satisfied in practice with a reasonably high probability, enabling the results in this paper on the optimality of treating interference as noise to be put into use in practice.

\section{Conclusion and Future Directions}
We introduced a condition on $K$-user interference channels under which power control at the transmitters and treating interference as noise at the receivers, in short, the TIN scheme, was proven to be GDoF-optimal. The GDoF region under this condition was shown to be a polyhedron. The analysis was also generalized to show that under the same condition, TIN can achieve the whole capacity region of the network to within a constant gap that only depends on the number of users $K$. Furthermore, the achievable GDoF region by TIN for general values of channel gains in a $K$-user interference channel was also characterized fully.

\begin{figure}
\begin{center}
 \includegraphics[width= 4.8 cm]{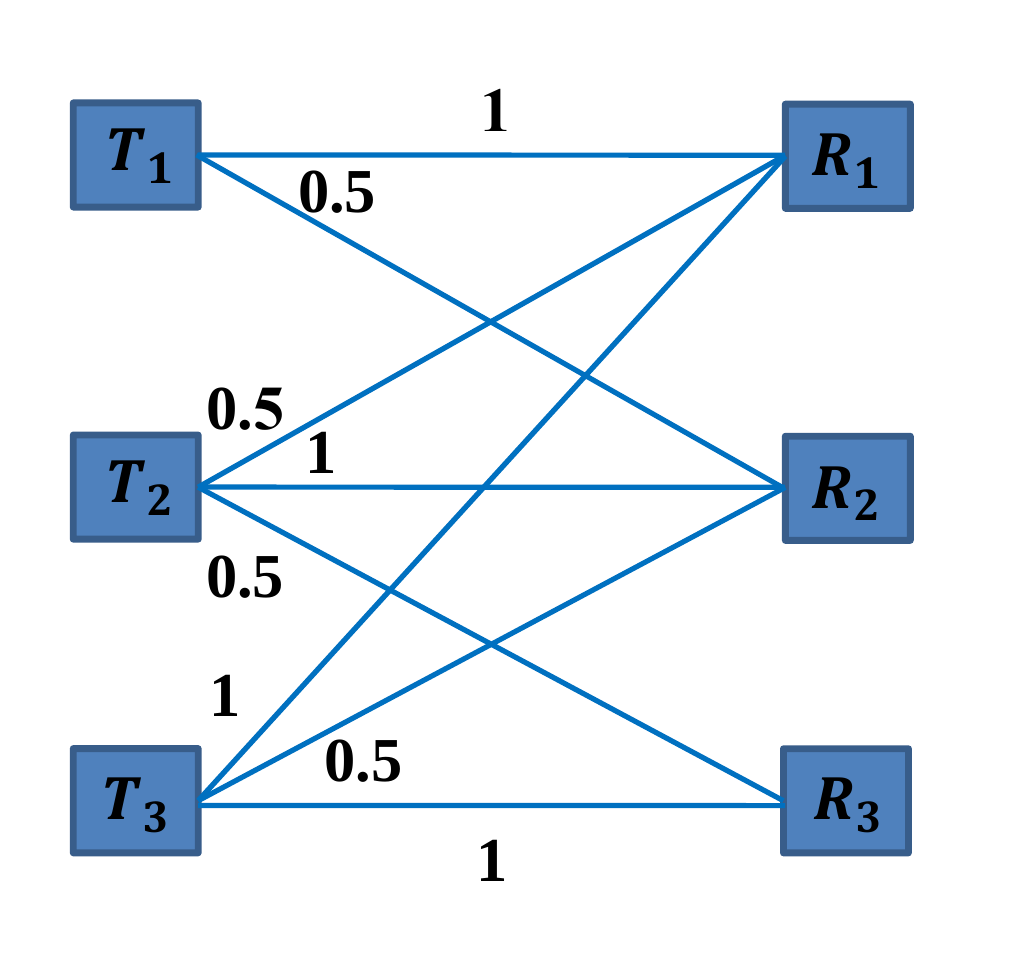}
 \caption{A 3-user network where (\ref{eqq10}) is not satisfied, but TIN is still optimal. The value on each link represents its channel strength level. All channel phases are assumed to be zero.}
\label{fig_conc}
\end{center}
\end{figure}

An interesting future direction is to determine whether the  condition (\ref{eqq10}) is also \emph{necessary} for TIN to be GDoF-optimal. For the 2-user case, it can be shown that except for a set of channel gain values with measure zero, the condition is also necessary for the GDoF-optimality of TIN. However, going beyond two users makes the problem more challenging. For instance, consider the network in Fig. \ref{fig_conc}. In this network,  condition (\ref{eqq10}) is violated at users 1 and 3. However, TIN is still optimal, because it is possible to show\footnote{Note that in this network, receiver 1, after decoding its own message and subtracting it from its received signal, has the same signal as receiver 3. Therefore, it is also able to decode the message of user 3. Besides, transmitters 1 and 3 have the same channel vectors to receivers 1 and 2. Therefore, the sum-GDoF of this network is upper bounded by the sum-GDoF of a 2-user interference channel with $\{\text{T}_{1,3},\text{T}_2\}$ as the transmitters and $\{\text{R}_1,\text{R}_2\}$ as the receivers, where $\text{T}_{1,3}$ is the combination of transmitters 1 and 3. In this 2-user interference channel, the sum-GDoF is equal to 1 \cite{Tse_GDoF}, therefore implying that the sum-GDoF of the network in Fig. \ref{fig_conc} is upper bounded by 1.} that the sum-GDoF is upper bounded by 1; i.e.,
\begin{align*}
d_1+d_2+d_3\leq 1.
\end{align*}

Moreover, Theorem \ref{th2} implies that the above region is in fact the polyhedral TIN region, hence showing that TIN is optimal in the network of Fig. \ref{fig_conc}, despite the fact that (\ref{eqq10}) does not hold in this network. Perturbing the channel strength levels from the values in Fig. \ref{fig_conc}, or even changing the phase values if they are available to the transmitters, will invalidate the converse. We  suspect that again outside a set of measure zero,  condition (\ref{eqq10}) is necessary for TIN to be GDoF-optimal. However, the necessity of this condition for the networks comprising more than 2 users remains open.

Another interesting direction following this work could be to extend the TIN scheme to include Gaussian superposition coding at the transmitters and successive interference cancellation at the receivers. In particular, it would be valuable to determine, with this additional flexibility, how much gain one can  obtain beyond TIN for general channel strength levels in a $K$-user interference channel. For the case of 2-user interference channels, this problem is studied in \cite{2userSIC}, and the regimes under which such schemes outperform TIN are identified.  Also, for the case of  $K$-user linear deterministic interference channels, the achievable rates of such schemes are characterized as the convex hull of the feasible rates supported by the independent sets  of an extended conflict graph~\cite{KuserDet}. However, general conditions for the optimality  of superposition coding and successive interference cancellation are still unknown.

\section*{Appendix: Proof of Theorem \ref{General_TIN_Region}}\label{TIN_region_appendix}
We prove the theorem in two steps.
\begin{itemize}
\item Step 1: $\bigcup_{\mathcal{S}\subseteq \{1,...,K\}} \mathcal{P}_\mathcal{S}\subseteq \mathcal{P}^*$. It suffices to show that for all $\mathcal{S}\subset \{1,...,K\}$, $\mathcal{P}_\mathcal{S}\subseteq \mathcal{P}^*$; i.e., the region $\mathcal{P}_\mathcal{S}$ can be achieved through TIN. Note that if $\mathcal{S}=\emptyset$, then $\mathcal{P}_\mathcal{S}=\mathcal{P}_\emptyset=\mathcal{P}\subseteq \mathcal{P}^*$.

Now, if $\mathcal{S}\neq\emptyset$, then to make the users in $\mathcal{S}$ silent, we set $r_i=-\infty$, $\forall i\in\mathcal{S}$. This forces $d_i=0$, $\forall i\in\mathcal{S}$. Then, for the remaining users, i.e., the users in $\mathcal{S}^c$, we use polyhedral TIN. Therefore, the polyhedral TIN region where all the users in $\mathcal{S}$ are removed from the network, can be achieved. This region is in fact $\mathcal{P}_\mathcal{S}$, and hence, $\mathcal{P}_\mathcal{S}\subseteq \mathcal{P}^*$.

\item Step 2: $\mathcal{P}^*\subseteq \bigcup_{\mathcal{S}\subseteq \{1,...,K\}} \mathcal{P}_\mathcal{S}$. To prove this, we first define the sets $\tilde{\mathcal{P}}_\mathcal{S}$ as $\mathcal{P}_\mathcal{S}$ restricted to strictly positive GDoF's for users in $\mathcal{S}^c$; i.e.,
\begin{align*}
\tilde{\mathcal{P}}_\mathcal{S}=\{(d_1,...,d_K)\in \mathcal{P}_\mathcal{S}: d_i>0, \forall i\in \mathcal{S}^c \},
\end{align*}
for any $\mathcal{S}\subseteq\{1,2,...,K\}$. It is obvious that $\tilde{\mathcal{P}}_\mathcal{S}\subseteq\mathcal{P}_\mathcal{S}$ and therefore,
\begin{equation}\label{PPtilde}
\bigcup_{\mathcal{S}\subseteq \{1,...,K\}} \tilde{\mathcal{P}}_\mathcal{S}\subseteq\bigcup_{\mathcal{S}\subseteq \{1,...,K\}} \mathcal{P}_\mathcal{S}.
\end{equation}

Now, we show that any GDoF point $(d_1,...,d_K)$ lying outside all of the sets $\tilde{\mathcal{P}}_\mathcal{S}$ should not be achievable by TIN. Such a point should satisfy at least one of the following conditions:
\begin{itemize}
\item $d_i<0$ or $d_i>\alpha_{ii}$ for some user $i\in\{1,...,K\}$. In this case, it is trivial that the GDoF point is not achievable by TIN.

\item $\sum_{j=1}^m d_{i_j}> \sum_{j=1}^m (\alpha_{i_j i_j}-\alpha_{i_{j-1} i_{j}})$ for some cyclic sequence $(i_0,i_1,...,i_m)\in\Pi_K$ such that $d_{i_j}>0$, $\forall j\in\{1,...,m\}$. In this case, letting $i_{m+1}=i_1$, we have
\begin{align*}
\sum_{j=1}^m r_{i_j}+\alpha_{i_ji_j}-\max\{0,\max_{i_k\neq i_j} (r_{i_k}+\alpha_{i_j i_k})\} &> \sum_{j=1}^m (\alpha_{i_j i_j}-\alpha_{i_{j-1} i_{j}})\\
\Rightarrow \sum_{j=1}^m r_{i_j}+\alpha_{i_{j-1}i_{j}}-\underbrace{\max\{0,\max_{i_k\neq i_j} (r_{i_k}+\alpha_{i_j i_k})\}}_{\geq r_{i_{j+1}}+\alpha_{i_{j}i_{j+1}}} &> 0,
\end{align*}
which is a contradiction. Therefore in this case, the GDoF point is not achievable by TIN, too.
\end{itemize}
This implies that $\mathcal{P}^*\subseteq \bigcup_{\mathcal{S}\subseteq \{1,...,K\}} \tilde{\mathcal{P}}_\mathcal{S}$, and combining this with (\ref{PPtilde}) yields $\mathcal{P}^*\subseteq \bigcup_{\mathcal{S}\subseteq \{1,...,K\}} \mathcal{P}_\mathcal{S}$.
\end{itemize}

Combining steps 1 and steps 2 leads to (\ref{true_tin_region}), therefore completing the proof.\hfill \QED

\end{document}